\documentclass[twocolumn,showpacs,preprintnumbers,amsmath,amssymb]{revtex4}
\usepackage{graphics}
\usepackage{epsfig}
\usepackage{bm}

\begin{document}
\title{Phase diagram of Fe-doped Ni-Mn-Ga ferromagnetic shape-memory alloys}
\author{Daniel Soto, Francisco Alvarado Hern\'{a}ndez, Horacio Flores}
\affiliation{Centro de Investigaciones en Materiales Avanzados,
Miguel de Cervantes 120, Chihuahua, M\'{e}xico C. P. 31109.}

\author{Xavier Moya, Llu\'{i}s Ma\~nosa, Antoni Planes}
\email{toni@ecm.ub.es}

\affiliation{Departament d'Estructura i Constituents de la
Mat\`eria, Facultat de F\'isica, Universitat de Barcelona,
Diagonal 647, E-08028 Barcelona, Catalonia, Spain}

\author{Seda Aksoy, Mehmet Acet}

\affiliation{Fachbereich Physik, Experimentalphysik,
Universit\"{a}t Duisburg-Essen, D-47048 Duisburg, Germany}

\author{Thorsten Krenke}

\affiliation{ThyssenKrupp Electrical Steel GmbH, D-45881
Gelsenkirchen, Germany}

\date{\today}

\pacs{81.30.Kf, 75.50.Cc}

\begin{abstract}
We have studied the effect of Fe addition on the structural and
magnetic transitions in the magnetic shape memory alloy Ni-Mn-Ga
by substituting systematically each atomic species by Fe.
Calorimetric and AC susceptibility measurements have been carried
out in order to study the magnetic and structural transformation
properties. We find that the addition of Fe modifies the
structural and magnetic transformation temperatures. Magnetic
transition temperatures are displaced to higher values when Fe is
substituted into Ni-Mn-Ga, while martensitic and premartensitic
transformation temperatures shift to lower values. Moreover, it
has been found that the electron per atom concentration
essentially governs the phase stability in the quaternary system.
However, the observed scaling of transition temperatures with
$e/a$ differs from that reported in the related ternary system
Ni-Mn-Ga.
\end{abstract}

\maketitle

\section{Introduction}

Magnetic shape-memory alloys have drawn much attention in recent
years owing to their unique magnetomechanical properties such as
magnetic shape-memory \cite{OHandley98} and the magnetic
superelasticity \cite{Krenke07}. These properties are a
consequence of a strong coupling between magnetic and structural
degrees of freedom. The prototypical and first discovered magnetic
shape-memory material is the Heusler Ni$_2$MnGa \cite{Ullakko96}.
This alloy undergoes a complex multi-stage transformation process
from a high temperature paramagnetic cubic phase to a
ferromagnetic martensitic phase. At intermediate temperatures it
shows precursor tweed textures which may lock (via a first-order
phase transition) into a modulated premartensitic structure due to
the freezing of a specific phonon with a given wave vector. This
behavior appears to be related to low resistance against
distortions of the $\{110\}$ planes along the $\langle 1 \bar 1 0
\rangle$ directions and is evidenced by the features of the low
energy TA$_2$ acoustic phonon branch
\cite{Zheludev95,Zheludev96,Stuhr97,Manosa01} and the low value of
the elastic constant $C'$ \cite{Worgull96,Manosa97,Stipcich04}.
While these features are essentially inherent to the
high-temperature cubic structure, additional softening has been
shown to arise from the coupling between structural and magnetic
degrees of freedom \cite{Stuhr97,Manosa01}. Thus, it has been
suggested that the magnetostructural coupling is responsible for
the phonon condensation yielding the intermediate modulated
structure \cite{Planes97}. Nevertheless, the occurrence of a
premartensitic phase is not yet a well understood phenomenon, as
it only has been observed for a restricted number of magnetic
shape memory alloys within limited composition ranges. Actually,
the study of the structural (martensitic and premartensitic
transformations) and magnetic properties of Ni-Mn-Ga alloys is a
current topic of intense research
\cite{Barman05,Bhobe06,Ranjan06,Banik06,PerezLandazabal07,Ahuja07,Banik07}.

The effect of doping elements on the martensitic and magnetic
transformations in Ni-Mn-Ga alloys has received considerable
attention
\cite{Liu02,Khovailo03b,Kikuchi04,Koho04,Guo05,Glavatskyy06,Ohtsuka06}.
However, the lack of a systematic study makes it difficult to
compare directly the properties of different compounds. In the
present paper, we investigate the dependence of transition
temperatures (martensitic, intermediate and Curie) on the electron
concentration by analyzing the effect of substituting Ni, Mn and
Ga by Fe. In all cases, the reference system is the stoichiometric
Ni$_2$MnGa, which has a high temperature L2$_1$ structure
($Fm3m$). This structure can be viewed as four interpenetrating
fcc sublattices [in Wickoff notation, (4a)-1 is occupied by
Mn-atoms, (4b)-2 by Ga-atoms, and (8c) by Ni-atoms]. The total
magnetic moment is $\sim 4.1 \mu_B$ per formula unit and is
largely confined to the Mn-sites contributing with $3.5 \mu_B$.

\section{Experimental}

\begin{table}
\begin{ruledtabular}\caption{\label{Tab1} Compositions of the
Ni-Mn-Ga-Fe samples determined by EDX. Different specimens are
grouped into three distinct families, depending on the element
that is substituted by Fe (elements within parenthesis, first
column). The estimated error in the compositions is less than $\pm
0.3$ \%. Values of valence electron concentration per atom, $e/a$,
are also given.}
\begin{tabular}{lccccc}
Family & Ni & Mn & Ga & Fe & $e/a$\\
 & (at. \%) & (at. \%) & (at. \%) & (at. \%) & \\
\hline\\
(Ni,Fe) & 52.6 & 23.1 & 24.3 & 0 \footnote{Data extracted from reference \cite{Hu01}.} & 7.606\\
        & 51.3 & 22.8 & 24.5 & 1.4 & 7.573\\
        & 50.1 & 23.1 & 24.6 & 2.2 & 7.541\\
        & 49.3 & 23.1 & 24.5 & 3.1 & 7.530\\
        & 48.1 & 23.0 & 24.5 & 4.4 & 7.507\\
        & 47.0 & 23.1 & 24.6 & 5.3 & 7.479\\
(Mn,Fe) & 51.4 & 24.8 & 23.8 &  0 \footnote{Data extracted from reference \cite{Wu03}. Note that this composition slightly deviates (more than the experimental error, $\pm 0.3 \%$) from the fitted compositional line.} & 7.589\\
        & 51.5 & 24.2 & 23.5 & 0.8 & 7.613\\
        & 51.1 & 24.6 & 23.4 & 0.9 & 7.606\\
        & 51.7 & 23.1 & 23.4 & 1.8 & 7.633\\
(Ga,Fe) & 51.3 & 24.0 & 24.7 & 0 \footnote{Data extracted from reference \cite{Tickle99}. Note that this composition slightly deviates (more than the experimental error $\pm 0.3 \%$) from the fitted compositional line.}& 7.551\\
        & 51.2 & 24.2 & 23.8 & 0.8 & 7.592\\
        & 51.8 & 24.8 & 21.7 & 1.7 & 7.703\\
        & 51.3 & 24.5 & 22.2 & 2.0 & 7.671\\
\end{tabular}
\end{ruledtabular}
\end{table}

Polycrystalline Ni-Mn-Ga-Fe ingots were prepared by arc melting
pure metals under argon atmosphere in a water cooled Cu crucible.
The ingots were melted several times for homogeneity and
encapsulated under vacuum in quartz glass. They were then annealed
at 1073 K for 72 hours to achieve a high degree of atomic order.
Finally, the samples were quenched in ice-water. The compositions
of the alloys were determined by energy dispersive x-ray
photoluminescence analysis (EDX) with an estimated error less than
$\pm 0.3$\% (Table \ref{Tab1}). The alloys are grouped according
to their compositions into the families
Ni$_{52.5-x}$Mn$_{23}$Ga$_{24.5}$Fe$_{x}$ (1.2 $\leq x\leq$ 5.5)
for which Ni is replaced by Fe;
Ni$_{51.4}$Mn$_{25.2-x}$Ga$_{23.4}$Fe$_{x}$ (0.8 $\leq x\leq$ 1.8)
for which Mn is replaced by Fe; and
Ni$_{51.4}$Mn$_{24.5}$Ga$_{24.1-x}$Fe$_{x}$ (0.7 $\leq x\leq$ 2.0)
where Fe replaces Ga. The compositions are given in at\%.

\begin{figure}
\includegraphics[width=8cm]{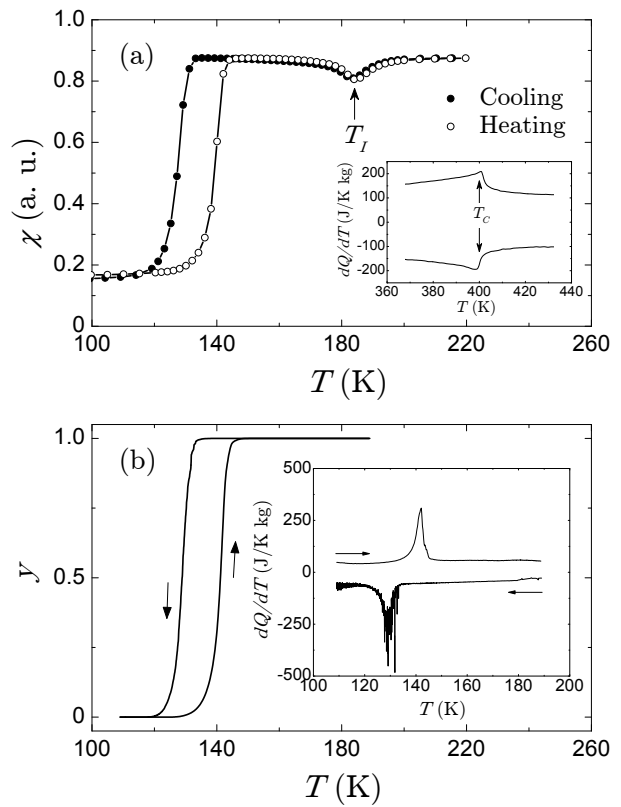}
\caption{\label{fig01} Ni$_{52.5-x}$Mn$_{23}$Ga$_{24.5}$Fe$_{x}$
family represented by the sample with $x=4.4$. (a) Magnetic
susceptibility versus temperature. The vertical arrow indicates
the premartensitic transition temperature, $T_{I}$. The inset
shows high temperature calorimetric curves. The Curie point
$T_{C}$ is indicated by vertical arrows. (b) Transformed fraction
as a function of temperature obtained by integration of the
calorimetric curves (inset in b). The arrows indicate the
direction of temperature change.}
\end{figure}

Specimens cut from the ingots using a low speed diamond saw
(typical size $5 \times 1 \times 1$ mm$^{3}$) were used as samples
for susceptibility and calorimetric studies. Structural transition
temperatures were obtained from AC susceptibility and calorimetric
measurements. Magnetic susceptibility measurements were carried
out in an AC susceptometer (LakeShore 7120A) in the temperature
range 80 K $\leq T\leq$ 320 K. The working parameters were 500 A
m$^{-1}$ (6.28 Oe) applied field and 389 Hz frequency. For
differential scanning calorimetry (DSC) measurements, one side of
the samples was ground with SiC abrasive to ensure optimal thermal
contact. Calorimetric measurements were carried out by means of a
high sensitivity calorimeter in the temperature range 100 K $\leq
T\leq$ 350 K. Typical heating and cooling rates were 0.5 K
min$^{-1}$. Magnetic transition temperatures were determined by
means of a DSC calorimeter suitable for higher temperatures. All
transition temperatures are affected by an error of $\pm 1$ K. The
errors in entropy change are based on reproducibility and shown as
errors bars in the figures.

\section{Experimental Results}

\begin{figure}
\includegraphics[width=8cm]{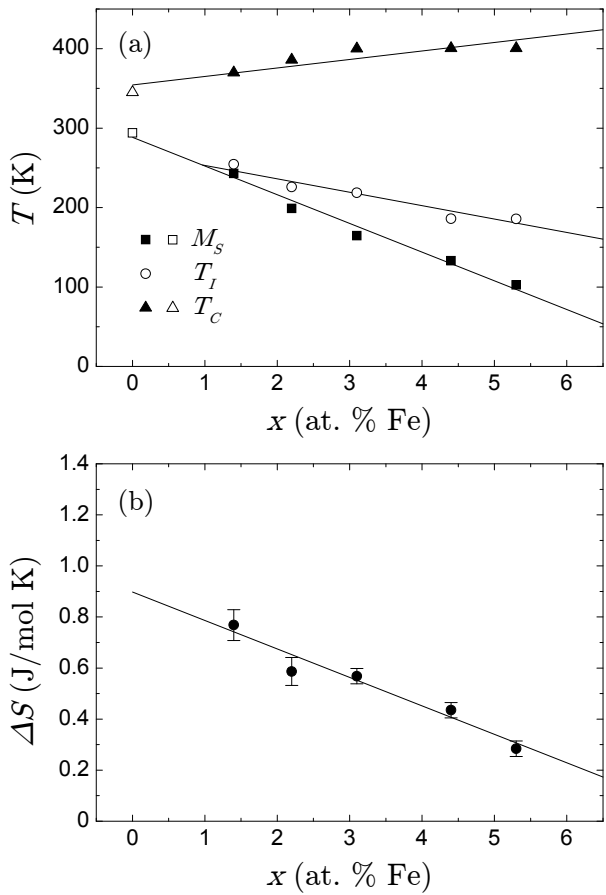}
\caption{\label{fig02} (a) Evolution of the transition
temperatures of Ni$_{52.5-x}$Mn$_{23}$Ga$_{24.5}$Fe$_{x}$ as a
function of Fe concentration. Open square and triangle symbols
stand for data extracted from ref. \cite{Hu01}. (b) Entropy change
at the martensitic transformation as a function of Fe
concentration. Solid lines are linear fits to the experimental
data.}
\end{figure}

Eleven different alloys were studied in the present work. In this
section, we present selected results of susceptibility and
calorimetric measurements which are representative of each family.
In the following the given Fe content is taken as the value
corresponding to the fitted compositional line. From the complete
set of data, we determine a phase diagram for each family and the
transition entropy change at the martensitic transformation.

\subsection{Substitution of Ni by Fe}

\begin{figure}
\includegraphics[width=8cm]{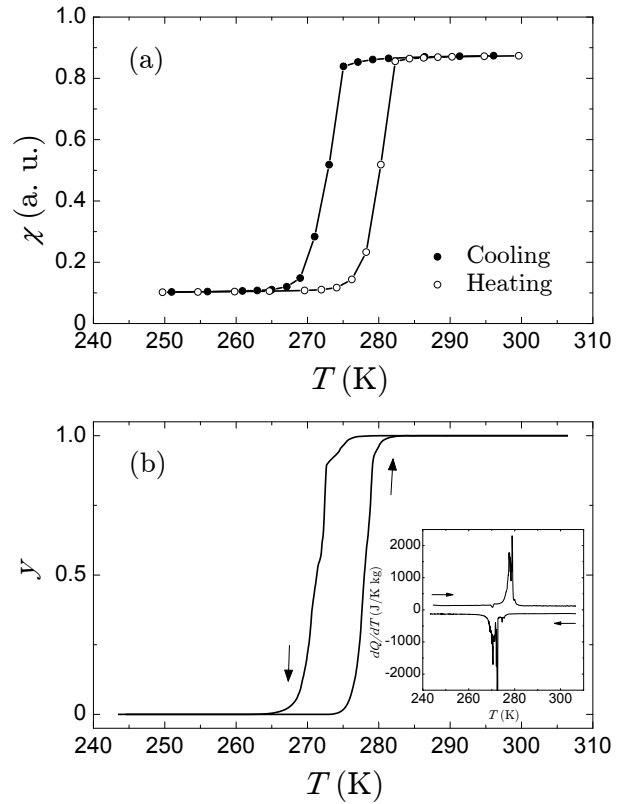}
\caption{\label{fig03} Ni$_{51.4}$Mn$_{25.2-x}$Ga$_{23.4}$Fe$_{x}$
family represented by the sample with $x=1.8$. (a) Magnetic
susceptibility versus temperature and (b) transformed fraction as
a function of temperature, obtained by integration of the
calorimetric curves (shown in the inset). Arrows in panel (b) and
inset indicate direction of temperature change.}
\end{figure}

Figure \ref{fig01} shows the AC susceptibility and calorimetric
curves for the sample with $x=4.4$. The inset in figure
\ref{fig01}(b) shows the calorimetric curves recorded on cooling
and heating. The multiple peaks (noticeable in the thermograms
corresponding to the forward transition on cooling) are a
consequence of the well-known jerky character of martensitic
transformations. On the other hand, the extra noise observed at
the lowest temperatures in the thermograms on cooling is an
artifact arising from the very low cooling rate in the low
temperature regime (notice that $dQ/dT$ is obtained by dividing
the calorimetric signal $\dot{Q}$ by $\dot{T}$). Figure
\ref{fig01}(b) shows the austenitic transformed fraction, $y$
versus $T$, obtained from the calorimetric data shown in the
inset. The austenitic transformed fraction is computed as
$y=1-\Delta S(T)/\Delta S$ for the forward transition on cooling,
and $y=\Delta S(T)/\Delta S$ for the reverse transition on
heating, with $\Delta S(T)=\int_{T_{i}}^{T} (dQ/dT)/T \, dT$
($T<T_{i}$ on cooling and $T>T_{i}$ on heating) and $\Delta S$,
the entropy change at the martensitic transformation. This plot is
illustrative for the typical results obtained for the
Ni$_{52.5-x}$Mn$_{23}$Ga$_{24.5}$Fe$_{x}$ family. Both
susceptibility and calorimetric measurements reveal the presence
of a martensitic transformation. The corresponding transition
temperatures are: martensite start temperature $M_{s}=133$ K,
martensite finish temperature $M_{f}=119$ K, austenite start
temperature $A_{s}=132$ K and austenite finish temperature
$A_{f}=146$ K. The Curie point was determined from complementary
DSC measurements as $T_{C}=400$ K [shown in the inset of Fig.
\ref{fig01}(a)]. Moreover, an additional feature is observed in
the susceptibility curve at temperatures above the martensitic
transition which is associated with the formation of the
intermediate or premartensitic phase \cite{Manosa97}. The
transition temperature is $T_{I}= 186$ K. No significant thermal
hysteresis is detected at the premartensitic transition and no
appreciable features are observed in the calorimetric curves at
the premartensitic transition. This behaviour agrees with that
observed in the related system Ni-Mn-Ga, where thermal anomalies
are barely detected with differential scanning calorimetric
techniques \cite{Kokorin96}. By contrast, AC susceptibility
measurements are very suited for the observation of the
intermediate phase transition \cite{Manosa97}.

Figure \ref{fig02}(a) summarizes the results for the
Ni$_{52.5-x}$Mn$_{23}$Ga$_{24.5}$Fe$_{x}$ family. To complete the
picture, we have also included data for an $x=0$ sample from
reference \cite{Hu01}. Transition temperatures are plotted as a
function of the Fe concentration. All transition temperatures
associated with the martensitic transformation ($M_{s}$, $M_{f}$,
$A_{s}$ and $A_{f}$) follow the same $x$ dependence. Thus, for the
sake of clarity, only $M_{s}$ temperatures are included. As can be
seen from this figure, the martensitic transformation temperature
decreases as the amount of Fe increases. In ternary Ni-Mn-$X$
($X$: Ga, Al, Sn, In and Sb) systems it is well established that
martensitic transformation temperatures decrease as the valence
electron concentration $e/a$ decreases
\cite{Chernenk99,Acet02,Krenke05,Krenke06}. When replacing Ni by
Fe, $e/a$ decreases and a drop in $M_s$ is expected. This behavior
is seen in Fig. \ref{fig02}(a).

\begin{figure}
\includegraphics[width=8cm]{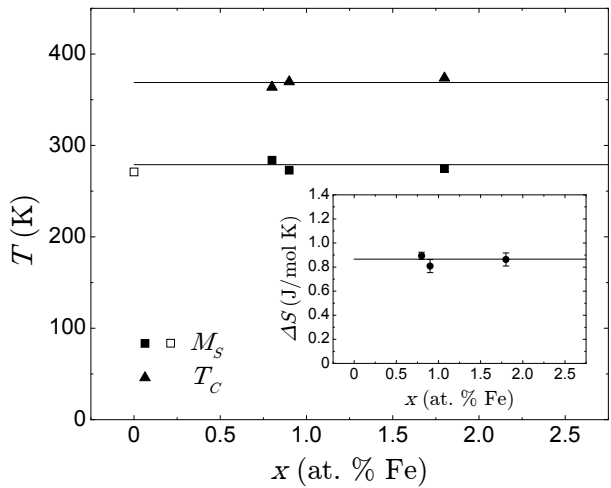}
\caption{\label{fig04} Transition temperatures for
Ni$_{51.4}$Mn$_{25.2-x}$Ga$_{23.4}$Fe$_{x}$ as a function of Fe
concentration. Open square symbol stands for data extracted from
ref. \cite{Wu03}, for this sample $T_{C}$ was not reported. The
inset shows the entropy change at the martensitic transformation
as a function of Fe concentration. Solid lines are fits to the
experimental data.}
\end{figure}

Premartensitic transformation temperatures also decrease as the Fe
concentration increases, but at lower rate than $M_s$. In addition
$T_C$ increases with increasing $x$.

Figure \ref{fig02}(b) shows the entropy change at the martensitic
transformation as a function of Fe concentration. The
concentration dependence of $\Delta S$ is similar to the behaviour
of $M_{s}$, i. e., the entropy change decreases as the amount of
Fe increases. Such a dependence reflects the stabilization of the
cubic phase.

\subsection{Substitution of Mn by Fe}

Figure \ref{fig03} illustrates typical results obtained when
replacing Mn by Fe (Ni$_{51.4}$Mn$_{25.2-x}$Ga$_{23.4}$Fe$_{x}$
family). For the sample with $x=1.8$ ($T_{C}=374$ K) a martensitic
transition is observed on cooling at $M_{s}=275$ K and $M_{f}=267$
K. On heating, the reverse transformation takes place at
$A_{s}=274$ K and $A_{f}=281$ K. No signatures of a premartensitic
transformation are observed.

\begin{figure}
\includegraphics[width=8cm]{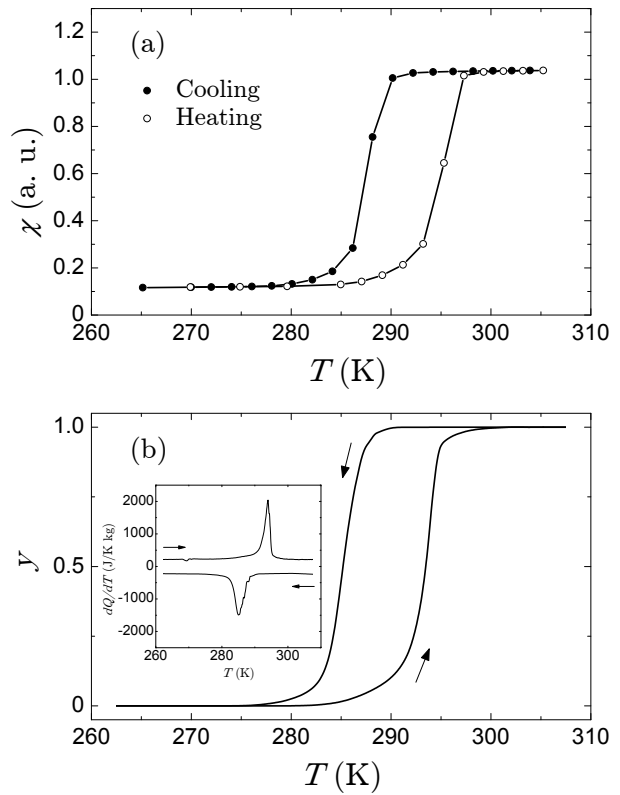}
\caption{\label{fig05} Ni$_{51.4}$Mn$_{24.5}$Ga$_{24.1-x}$Fe$_{x}$
family represented by the sample with $x=0.7$. (a) Magnetic
susceptibility versus temperature and (b) transformed fraction as
a function of temperature, obtained by integration of the
calorimetric curves (shown in the inset). Arrows in panel (b) and
inset indicate direction of temperature change.}
\end{figure}

The variation of transition temperatures with Fe concentration for
this family is collected in Fig. \ref{fig04}. No significant
changes in transition temperatures are observed over the
compositional range studied. This is because $e/a$ varies little
by replacing Mn with Fe in small amounts. Consistently, Fe
addition does not substantially modifies the values of the entropy
change at the martensitic transition, as can be seen in the inset
of figure \ref{fig04}.

\subsection{Substitution of Ga by Fe}

\begin{figure}
\includegraphics[width=8cm]{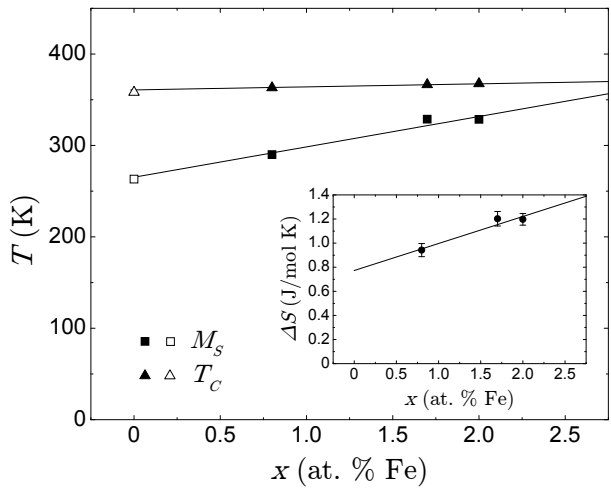}
\caption{\label{fig06} Transition temperatures for
Ni$_{51.4}$Mn$_{24.5}$Ga$_{24.1-x}$Fe$_{x}$ as a function of Fe
concentration. Open square and triangle symbols stand for data
extracted from ref. \cite{Tickle99}. The inset shows the entropy
change at the martensitic transformation as a function of Fe
concentration. Solid lines are linear fits to the experimental
data.}
\end{figure}

Figure \ref{fig05} illustrates typical results obtained for the
Ni$_{51.4}$Mn$_{24.5}$Ga$_{24.1-x}$Fe$_{x}$ family. Data for the
sample with $x=0.7$ ($T_{C}=363$ K) are shown. The presence of a
martensitic transformation near room temperature is evidenced from
both susceptibility and calorimetric measurements. The
corresponding transition temperatures are $M_{s}=290$ K,
$M_{f}=281$ K, $A_{s}=287$ K and $A_{f}=297$ K. Again, no
signature of the premartensitic transition is observed.

The phase diagram is shown in figure \ref{fig06}, where it is seen
that $M_s$ increases with increasing Fe content. This is
consistent with the rapid increase of $e/a$ when Fe is substituted
for Ga. $T_C$ is essentially unaffected.

The entropy change at the martensitic transition as a function of
Fe concentration is collected in the inset of figure \ref{fig06}.
As can be seen from this figure, $\Delta S$ parallels the
behaviour of the martensitic transformation temperatures and
increases as the amount of Fe increases, pointing out the
stabilization of the the low temperature phase due to Fe
substitution.

\begin{figure}
\includegraphics[width=8cm]{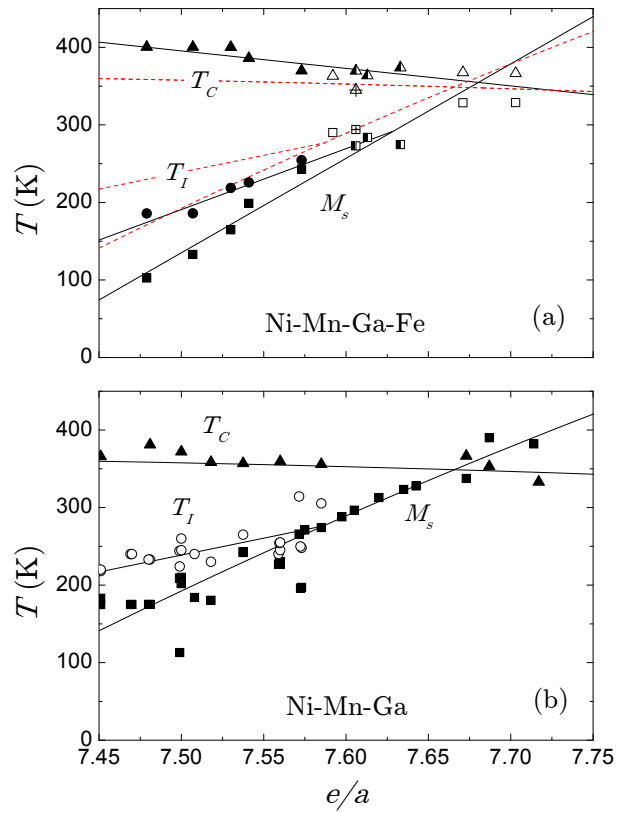}
\caption{\label{fig07} (Color online) (a) Phase diagram of
Ni-Mn-Ga-Fe system as a function of electron concentration per
atom $e/a$. Filled symbols stand for
Ni$_{52.5-x}$Mn$_{23}$Ga$_{24.5}$Fe$_{x}$ family; half-filled
symbols stand for Ni$_{51.4}$Mn$_{25.2-x}$Ga$_{23.4}$Fe$_{x}$
family; open symbols stand for
Ni$_{51.4}$Mn$_{24.5}$Ga$_{24.1-x}$Fe$_{x}$ family; crossed
symbols stand for data extracted from reference \cite{Hu01}. Red
dashed lines depict the (fitted) transition lines of the related
Ni-Mn-Ga ternary system. (b) Phase diagram of Ni-Mn-Ga system as a
function of electron per atom concentration $e/a$ (data compiled
from reference \cite{Marcos04}). Solid lines are fits to the
experimental data.}
\end{figure}

\section{Discussion}

The complete set of results for the different transition
temperatures is collected in Fig. \ref{fig07}. Here, the magnetic
and structural transition temperatures of the quaternary
Ni-Mn-Ga-Fe system is plotted as a function of $e/a$. As can be
seen from this plot, data from different families scale with the
electron concentration parameter. It was established for Ni-Mn-Ga
that the phase stability is controlled by $e/a$
\cite{Chernenk99,Jin02}. In the case of the quaternary system, the
reasonable scaling of the transition temperatures indicates that
the phase stability is mostly governed by the electron
concentration as well. However, the scatter in the data points is
higher than that observed in the phase diagram as a function of
composition (see Figs. \ref{fig02}, \ref{fig04} and \ref{fig06}),
thus suggesting that additional parameters other than electron
concentration could affect phase stability.

For comparison, figure \ref{fig07}(b) shows the phase diagram for
the Ni-Mn-Ga system (data extracted from reference \cite{Marcos04}
and references therein). The behavior is similar for both alloy
systems. $M_{s}$ and $T_{I}$ increase as $e/a$ increases, whereas
$T_{C}$ decreases. At constant $e/a$, we find that the addition of
Fe to Ni-Mn-Ga shifts $M_s$ and $T_I$ to lower values, whereas
$T_C$ shifts to higher temperatures [as illustrated by dashed
lines in Fig. \ref{fig07}(a)].

The relationship of $e/a$ and lattice instability in cubic Heusler
alloys has recently been investigated from first principles
calculations \cite{Zayak06}. It has been reported that $e/a$ plays
a central role in the occurrence of anomalies in the phonon
dispersion curves along [110] directions. These control the
stability of the cubic structure. In particular, it has been found
that adding and removing electrons has the same effect as
replacing the $sp$ ($X$) element. In the present study, we have
experimentally investigated the effect of different element
substitution. The general trends in the phase stability are given
by the change in $e/a$. This is consistent with a change in the
position of the Fermi energy as in a rigid band model.
Nevertheless, the larger scatter of the data when plotted as a
function of $e/a$ compared to the one in the plots as a function
of composition suggests that the effect of alloying is not just a
change in the Fermi level, but the addition of Fe could also
modify to some extent the orbital hybridization and bonding.
Actually, changes in hybridization were reported for Ni$_2$MnGa
with several substitutional elements \cite{MacLaren02}. This could
be related to volume effects which have been reported for In-doped
Ni-Mn-Ga alloys \cite{Khan04}.

\begin{figure}
\includegraphics[width=8cm]{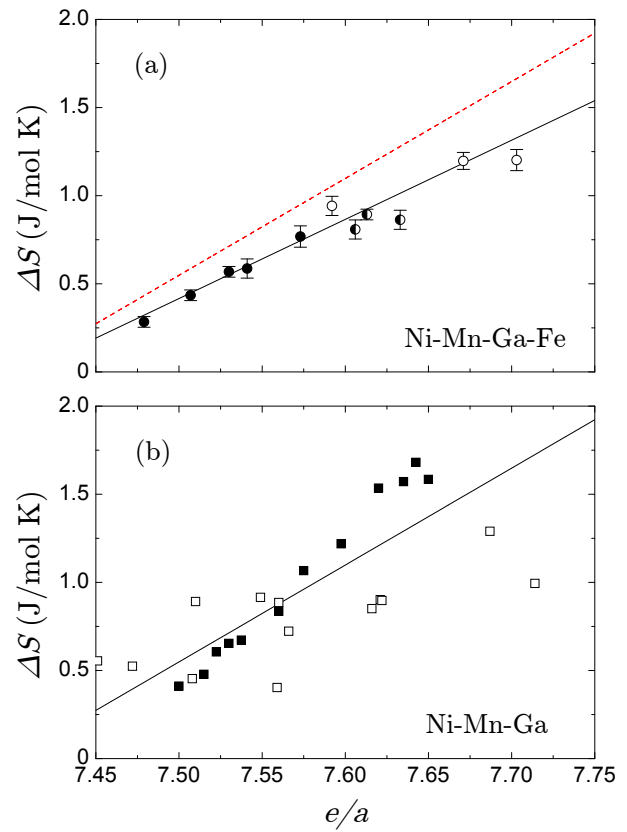}
\caption{\label{fig08} (Color online) (a) Entropy change at the
martensitic transformation of Ni-Mn-Ga-Fe system as a function of
electron concentration per atom $e/a$. Filled symbols stand for
Ni$_{52.5-x}$Mn$_{23}$Ga$_{24.5}$Fe$_{x}$ family; half-filled
symbols stand for Ni$_{51.4}$Mn$_{25.2-x}$Ga$_{23.4}$Fe$_{x}$
family; open symbols stand for
Ni$_{51.4}$Mn$_{24.5}$Ga$_{24.1-x}$Fe$_{x}$ family. Red dashed
line depicts the (fitted) entropy change of the related Ni-Mn-Ga
ternary system. (b) Entropy change at the martensitic
transformation of Ni-Mn-Ga system as a function of electron per
atom concentration $e/a$ [data compiled from reference
\cite{Chernenko95} ($\square$) and \cite{Khovailo03}
($\blacksquare$)]. Solid lines are linear fits to the experimental
data.}
\end{figure}

As can be seen in Fig. \ref{fig07}, the premartensitic phase
exists when martensitic and magnetic transition are well
separated. In the Ni-Mn-Ga system, it has been shown that
magnetoelastic coupling between structural and magnetic degrees of
freedom gives rise to the premartensitic transition
\cite{Planes97,Castan99}. The strength of such an interaction
depends on the magnetization. Therefore, in order for the
premartensitic phase to develop, the sample must remain in the
cubic phase at temperatures well below the Curie point. This
requires that the martensitic transition temperature is well below
$T_{C}$. Moreover, the temperature that corresponds to the point
where martensitic and premartensitic transformation temperatures
meet is slightly displaced to higher $e/a$ values in the case of
Ni-Mn-Ga-Fe system with respect to the ternary system. Such a
shift is in agreement with the decrease of $M_s$ and the increase
of $T_C$ due to Fe addition. As $M_s$ shifts to lower temperatures
and $T_C$ to higher temperatures, the separation between both
temperatures increases compared to the ternary system for equal
$e/a$ values. Thus, the crossing point between $M_{s}$ and $T_{I}$
is displaced to higher electron concentration values.

The features in the [110] TA$_2$ phonon branch giving rise to the
intermediate phase are associated with a nesting in the Fermi
surface. It has been found that such a Fermi-surface nesting is
strongly dependent on the magnetization of the cubic phase
\cite{Lee02}. This scenario is consistent with the experimental
finding that the premartensitic phase only develops for
ferromagnetically ordered samples for which the martensitic
instability is well below $T_C$.

Finally, figure \ref{fig08} shows the entropy change at the
martensitic transformation as a function of electron concentration
per atom $e/a$ for (a) Ni-Mn-Ga-Fe and (b) Ni-Mn-Ga systems. As
can be seen from panel (a), in the quaternary system $\Delta S$
increases as the electron per atom concentration increases,
similar to the behaviour exhibited by the martensitic
transformation temperatures and to the behaviour of the ternary
system. Moreover, the entropy change values in the Fe substituted
alloys are lower than those in the ternary Ni-Mn-Ga system, as
illustrated by the red dashed line. This drop could be accounted
for by the strengthening of magnetic exchange interactions when
adding Fe, as reflects the increase of $T_{C}$ in the quaternary
system compared to the ternary one. When magnetic order occurs in
the parent phase, the Gibb's chemical free energy decreases,
compared to the non magnetic state. Thus, the difference in the
free energy between parent and martensite phases is smaller and
the parent phase becomes more stable. Actually, the magnetic
contribution is also responsible of the strong concentration
dependence of the entropy change, as was pointed out by Khovailo
\textit{et al.} in the ternary Ni-Mn-Ga system \cite{Khovailo03}.

\section{Conclusion}

We have studied the effect of Fe addition on the structural and
magnetic transformation properties in the magnetic shape memory
alloy Ni-Mn-Ga for compositions close to stoichiometry. We find
that $M_s$ and $T_I$ shift to lower values when Fe is substituted
into Ni-Mn-Ga, while $T_C$ shift to higher values. Despite of the
similarities between ternary Ni-Mn-Ga and quaternary Ni-Mn-Ga-Fe
systems, which indicate that phase stability is qualitatively
governed by $e/a$, the shift in $M_s$ evidences that parameters
other than $e/a$ affect phase stability (essentially volume
effects associated with atom sizes as suggested in \cite{Khan04}).
Hence, a simple choice of $e/a$ can only be considered to be a
guideline for examining systematic changes within a single-alloy
system. Actually, the lack of universal character of $e/a$
parameterization has been previously pointed out for the Heusler
alloys Ni-Mn-$X$ \cite{Krenke06,KrenkeJMMM07} and has been
recently confirmed by the manipulation of structural and magnetic
transition temperatures in isoelectronic Ni-Mn-Ga and Ni-Mn-Ga-In
compounds \cite{Khan04,Aksoy07}.

\begin{acknowledgments}
This work received financial support from the CICyT (Spain),
Project No. MAT2007--61200,  DURSI (Catalonia), Project No.
2005SGR00969, from the Deutsche Forschungsgemeinschaft (GK277),
from Marie-Curie RTN Multimat (Contract No. MRTN-CT-2004-505226),
and from CONACYT (44786-SEP-CONACYT 2003). XM acknowledges support
from DGICyT (Spain). We thank Peter Hinkel for technical support.
\end{acknowledgments}


\end{document}